\documentclass[pra,twocolumn,showpacs,amsmath,amssymb]{revtex4}
\usepackage{amssymb}
\usepackage{dcolumn}
\usepackage{bm}
\usepackage{graphicx}
\usepackage{amsmath}

\begin{document}

\title{Photon-induced sideband transitions in a many-body Landau-Zener process}

\author{Honghua Zhong$^{1,2}$}
\author{Qiongtao Xie$^{1,3}$}
\author{Jiahao Huang$^{1}$}
\author{Xizhou Qin$^{1}$}
\author{Haiming Deng$^{1}$}
\author{Jun Xu$^{1}$}
\author{Chaohong Lee$^{1,}$}
\altaffiliation{Corresponding author. Email: chleecn@gmail.com}

\affiliation{$^{1}$State Key Laboratory of Optoelectronic Materials and Technologies, School of Physics and Engineering, Sun Yat-Sen University, Guangzhou 510275, China}

\affiliation{$^{2}$Department of Physics, Jishou University, Jishou 416000, China}

\affiliation{$^{3}$School of Physics and Electronic Engineering, Hainan Normal University, Haikou 571158, China}

\begin{abstract}

We investigate the many-body Landau-Zener (LZ) process in a two-site Bose-Hubbard model driven by a time-periodic field. We find that the driving field may induce sideband transitions in addition to the main LZ transitions. These photon-induced sideband transitions are a signature of the photon-assisted tunneling in our many-body LZ process. In the strong interaction regime, we develop an analytical theory for understanding the sideband transitions, which is confirmed by our numerical simulation. Furthermore, we discuss the quantization of the driving field. In the effective model of the quantized driving field, the sideband transitions can be understood as the LZ transitions between states of different ``photon" numbers.

\pacs{67.85.-d, 03.75.Lm, 03.75.Kk, 05.30.Jp}
\end{abstract}

\maketitle

\section{Introduction}

The Landau-Zener (LZ) model has long been a physical
paradigm for the tunneling process in a two-level
quantum  system subject to  a linearly time-dependent external field
\cite{Landau,Zener,Stueckelberg,Majorana}. Despite its simplicity,
this model  has been applied to a great variety of physical systems,
including atomic and molecular systems \cite{mgong,longhi},
semiconductor superlattices \cite{betthau}, and superconducting
devices \cite{william}. In these systems, the two-level LZ model
acts as  a good starting point to investigate more realistic
situations.

In recent years,  ultracold atomic  gases  have provided new
opportunities for investigating a many-body extension  of the simple
two-level LZ tunneling process \cite{Wu,Zobay,Liu,Witthaut,
Tomadin, Lee, Smith, korsch,lignier,Kasztelan}. The
remarkable controllability in these systems allows a clear study of
the effect of the interatomic interaction on the  LZ tunneling
process. For weakly interacting ultracold atomic  gases in optical
lattices,  the LZ tunneling between the two lowest Bloch bands has
been observed experimentally \cite{Arimondo,Salger, Kling}. It is
shown that due to the presence of the interatomic interaction, the
LZ tunneling can be enhanced if the system is initially in the
ground Bloch band, and be suppressed if the system is initially in
the higher Bloch band. In addition,  the many-body LZ
tunneling in the Mott-insulating regime has been addressed in
experiments \cite{Chena}.

Periodic driving fields have been extensively used to control quantum tunneling and transport~\cite{grifoni98}. Along this line, periodic driving fields have also been used to control LZ processes. For the simple two-level  LZ process, the effect of an additional periodic driving field in the bias or the coupling has been discussed~\cite{Kayanuma,Ful,malossi,wubs}. The driving field can induce interesting quantum-interference effects between two well-separated LZ sub-processes, and the probability of LZ transitions depends sensitively on the parameter values of the driving field.  Recently, the problem of how periodic driving field affect the nonlinear two-level LZ processes has been investigated~\cite{qzhang,qzhang08}.  Dependent on the nonlinearity strength,  the final transition probability shows a shifted phase-dependence on the driving field.  However, for the LZ process in an interacting many-body quantum system, the effects of the periodic driving field is still unclear.

In the present paper, we use a two-site Bose-Hubbard model to study
the effect of the periodic driving field on the many-body LZ process. In this
model,  the energy bias is subject to the usual linear change with
time superimposed by a time-periodic driving field.  We find that the
periodic driving field can modify the condition for the occurrence of
the many-body LZ tunneling. In addition to the original LZ transitions
without periodic driving field, sideband LZ transitions are induced. In
the high-frequency limit and strong interaction regime, we
obtain an effective system without periodic driving field to understand
these photon-assisted LZ transitions. The parametric dependence of
this photon-assisted LZ tunneling is analyzed. In addition, by quantizing the periodic driving field, we introduce the fully quantum mechanical model to understand the
sideband LZ transitions. Our results show
that the photon-assisted tunneling due to the periodic driving field
can provide an efficient way to control the many-body LZ tunneling process.

The structure of this article is as following. In section II, we give a physical description of  the two-mode Bose-Hubbard model where the energy bias is subject to a linear change with time and a time-periodic driving field. In section III, we discuss the effect of the periodic driving on the many-body LZ process  in the high-frequency limit and the strong interaction regime. In section IV, we discuss the quantization of the periodic driving field. In the last section, we briefly summarize our results.

\section{Model of many-body LZ processes in a diagonal periodic driving field}

\begin{figure}[htb]
\begin{center}
\includegraphics[bb= 240 160 503 393, clip, scale=0.5]{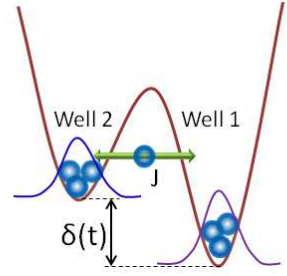}
\caption{(Color online) Schematic diagram for BEC in a double-well potential where the
time-dependent bias $\delta(t)$ is induced by an external field.} \label{fig0}
\end{center}
\end{figure}

The system under consideration is a gaseous BEC of bosonic atoms in a double-well potential \cite{smerzi97,corney,smerzi, martin,gati2007,Lee2012}, see its schematic diagrams in Fig. \ref{fig0}. The double-well potential is driven by external field which is composed of both a linearly time-dependent change with the sweep rate $S_0$ and a time-periodic
driving field with the amplitude $S_1$ and the frequency $\omega$, $S(t)=S_0t+S_1\cos (\omega t)$ \cite{Kayanuma,Ful,malossi}. Then the total system is described by a second quantized
Hamiltonian
\begin{eqnarray}
H(t)&=&H_{0}+H_{int}.\nonumber
\end{eqnarray}
The Hamiltonian
$H(t)$ includes two parts. $H_0$ is non-interacting part
\begin{eqnarray}
H_{0}=\int
\hat{\Psi}^{+}(x)[h_{0}+ S(t)V_{1}(x)]
\hat{\Psi}(x)dx, \label{hamiltonian}
\end{eqnarray}
with
\begin{eqnarray}
h_{0}=-\frac{\hbar^2\nabla^2}{2m_s}
+V_{0}(x).  \nonumber
\end{eqnarray}
Here $\hat{\Psi}^{(+)}(x)$ are the bosonic field operators
which annihilate (create) a particle at position $x$, and $m_s$ is the single-atom mass. $V_{0}(x)$ is of a symmetric double-well structure, and $V_{1}(-x)=-V_{1}(x)$ is a anti-symmetric, which can be realized in optical double-well experiment \cite{della,kartashov}. The Hamiltonian $H_{int}$ describes the two-body collisions between atoms, and is given by
\begin{eqnarray}
H_{int}=\frac{1}{2}g\int\hat{\Psi}^{+}(x)
\hat{\Psi}^{+}(x)\hat{\Psi}(x)
\hat{\Psi}(x)dx. \label{interh}
\end{eqnarray}
Here $g=4\pi\hbar^2a_{s}/m_s$ measures the interaction
strength between atoms, where
$a_{s}$ is the corresponding s-wave scattering length.  If the depth of the  symmetric double-well $V_0(x)$ is  enough
large, so that the dynamics is only involved in the two lowest
states localized to each well, we can apply the standard two-mode
approximation \cite{smerzi,martin}
\begin{equation}
\hat{\Psi}(x)=\hat{a}_1u_1(x)+\hat{a}_2u_2(x), \label{tmap}
\end{equation}
where $\hat{a}_j^{(\dagger)}\;(j=1,2)$ are the atomic annihilation (creation) operators for the
$i$-th well, $u_1(x)=[\phi_g(x)+\phi_e(x)]/\sqrt{2}$ and $u_2(x)=[\phi_g(x)-\phi_e(x)]/\sqrt{2}$ are localized waves in the wells 1 and 2, in which $\phi_{e}(x)$ and $\phi_{g}(x)$ are the two lowest energy eigenstates of $h_0$,  $h_0\phi_{e,g}(x)=E_{e,g}\phi_{e,g}(x)$. The total number of
atoms $N$ corresponding to the atom number operator $\hat{N}=\hat{a}_1^{\dagger}\hat{a}_1+\hat{a}_2^{\dagger}\hat{a}_2$ is
a conserved quantity.
This two-mode
approximation eventually simplifies Eq. (\ref{hamiltonian}) to
\begin{eqnarray}
H_0&=&-J(\hat{a}_{1}^\dagger \hat{a}_{2}+\hat{a}_{2}^\dagger
\hat{a}_{1})\nonumber \\
&&+\frac{1}{2}[\alpha t+\delta_1\cos(\omega t)](\hat{a}_{2}^{\dagger}\hat{a}_{2}
-\hat{a}_{1}^{\dagger}\hat{a}_{1}),
\end{eqnarray}
with the parameters
\begin{eqnarray}
J&=&-\int dx
[u_1^\ast(x)h_0u_{2}(x)], \nonumber \\
\alpha&=&2S_{0}\int dx
[u_1^\ast(x)V_{1}(x)u_{1}(x)], \nonumber \\
\delta_1&=&2S_{1}\int dx
[u_1^\ast(x)V_{1}(x)u_{1}(x)]. \nonumber
\end{eqnarray}
The term proportional to $J$ describes tunneling of particles from one to the other well, and we have assumed it to be real. In the same way, we stipulate that the overlap of
the functions $u_1(x)$ and $u_2(x)$ be only minute, which implies
that the condensates in the two wells are merely weakly
coupled, and ignore the high-order overlaps between two functions \cite{martin,gati2007,Lee2012}. Then the interaction between atoms $H_{int}$ is given as
\begin{eqnarray}
H_{int}=\frac{U_{11}}{4}\hat{a}_{1}^{\dagger}\hat{a}_{1}^{\dagger}\hat{a}_{1}\hat{a}_{1}+
\frac{U_{22}}{4}\hat{a}_{2}^{\dagger}\hat{a}_{2}^{\dagger}\hat{a}_{2}\hat{a}_{2},
\end{eqnarray}
with
\begin{eqnarray}
U_{jj}=2g\int dx
|u_{j}(x)|^4, \ \ j=1,2. \nonumber
\end{eqnarray}
After omitting the constant terms $O(N)$ and $O(N^2)$, the total Hamiltonian now can be rewritten as a two-mode Bose-Hubbard Hamiltonian
\begin{eqnarray}
H(t)&=&-J(\hat{a}_{1}^\dagger \hat{a}_{2}+\hat{a}_{2}^\dagger
\hat{a}_{1})+\frac{E_c}{8}(\hat{a}_{2}^{\dagger}\hat{a}_{2}
-\hat{a}_{1}^{\dagger}\hat{a}_{1})^2\nonumber
\\&+&\frac{\delta(t)}{2}(\hat{a}_{2}^{\dagger}\hat{a}_{2}
-\hat{a}_{1}^{\dagger}\hat{a}_{1}), \label{bhh}
\end{eqnarray}
with $E_c=U_{11}+U_{22}$ and
time-dependent energy bias
\begin{eqnarray}
\delta(t)=\alpha t+\delta_1\cos(\omega t).
\end{eqnarray}
In addition, by introducing the angular momentum
operators $S_{x}=(\hat{a}_{1}^\dagger \hat{a}_{2}+\hat{a}_{1}\hat{a}_{2}^\dagger)/2$,
$S_{y}=(\hat{a}_{2}^\dagger \hat{a}_{1}-\hat{a}_{1}^\dagger \hat{a}_{2})/2i$, and
$S_{z}=(\hat{a}_{2}^\dagger \hat{a}_{2}-\hat{a}_{1}^\dagger \hat{a}_{1})/2$ with the
Casimir invariant
\begin{figure}[htb]
\begin{center}
\includegraphics[bb= 160 85 538 520, clip, scale=0.5]{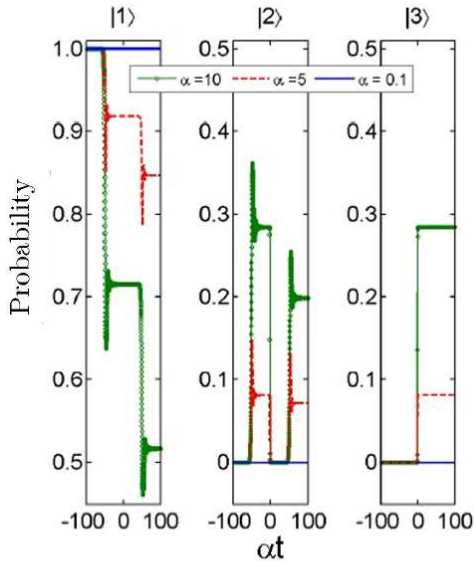}
\caption{(Color online) Occupation probability of the system  in the instantaneous
eigenstates of the Hamiltonian $H(t)_{\delta_1=0}$
 for three different values of the sweep rate $\alpha$  with $N=2$,  $J=1$, $E_c=100$, and $\delta_1=0$.  The system
initially starts  from its instantaneous ground state with  a large
negative bias. Here the solid lines are for $\alpha=0.1$, the dashed
lines are for $\alpha=5$, and the solid lines with circles are for
$\alpha=10$.  Here the labels $|1\rangle, |2\rangle$, and
$|3\rangle$ represent the three lowest instantaneous eigenstates of
the Hamiltonian $H(t)_{\delta_1=0}$.} \label{fig1}
\end{center}
\end{figure}
$S^{2}=(N/2)(N/2+1)$,
the Hamiltonian (\ref{bhh}) also can be
rewritten as
\begin{eqnarray} \label{Halmitquation}
H(t)=-2J S_{x}+\frac{E_c}{2} S_{z}^{2} +
\delta(t)S_{z}.
\end{eqnarray}
Clearly, the problem  with
$\delta_1=0$ is reduced to the usual many-body LZ problem
\cite{korsch}. We note that in the mean-field approximation where the system can be described by a nonlinear two-level model, the nonlinear LZ process  in a periodic driving field has been studied \cite{qzhang}.

\section{Photon-induced sideband transitions}

In the following, we discuss the effect of the additional
periodic driving on the many-body LZ tunneling process. By
expanding the state vector $|\psi(t)\rangle$ as a liner combination
of the Fock states $|N/2-n,N/2+n\rangle$, denoting the state with
$N/2-n$ bosons on the first well and $N/2+n$ on the second well,
\begin{eqnarray}
|\psi(t)\rangle=\sum_{n=-N/2}^{N/2}\exp[-i\theta_n(t)]
c_n(t)|N/2-n,N/2+n\rangle, \nonumber
\end{eqnarray}
with
\begin{eqnarray}
\theta_n(t)&=&\int_0^t[E_c n^2/2+(\alpha\tau+\delta_1\cos(\omega\tau)) n]d\tau \nonumber \\
&=&E_c n^2t/2+\alpha n t^2/2+\delta_1n\sin( \omega t)/\omega,
\nonumber
\end{eqnarray}
from the  time-dependent Schr\"{o}dinger equation $i\partial
|\psi(t)\rangle/\partial t =H(t)|\psi(t)\rangle$, we can get $N+1$
coupled first-order differential equations for the coefficients
$c_{n}(t)$
\begin{eqnarray} \label{dequation}
i\frac{dc_{n}}{dt}=&&-J\sqrt{(\frac{N}{2}-n)(\frac{N}{2}+n+1)}e^{-i\Delta\theta_{n+1}^{n}}c_{n+1} \nonumber \\
&&-J\sqrt{(\frac{N}{2}+n)(\frac{N}{2}-n+1)}e^{i\Delta\theta_{n}^{n-1}}c_{n-1},
\end{eqnarray}
with $\Delta\theta_{n+1}^{n}=\theta_{n+1}-\theta_n=E_c
(2n+1)t/2+\alpha t^2/2+\delta_1 \sin (\omega t)/\omega$. By applying
the generating function of the Bessel functions,  $\exp[\pm
i\delta_1\sin (\omega t)/\omega ]=\sum_{m=-\infty }^{\infty } J_{m}(
\delta_1/\omega) \exp [\pm i m\omega t]$, we have
\begin{eqnarray}\label{dequationb}
i\frac{dc_{n}}{dt}=&&-\sum_{m=-\infty }^{\infty }\widetilde{J}_{n,eff}^me^{-i\varphi_{n}^{m}}c_{n+1} \nonumber \\
&&-\sum_{m=-\infty }^{\infty
}\widetilde{J}_{n-1,eff}^{m}e^{i\varphi_{n-1}^{m}}c_{n-1},
\end{eqnarray}
with
\begin{eqnarray}
\widetilde{J}_{n,eff}^m&=&J\sqrt{(\frac{N}{2}-n)(\frac{N}{2}+n+1)}J_m(\delta_1/\omega),\\
\varphi_{n}^{m}&=&E_c (2n+1)t/2+\alpha t^2/2+m\omega t.
\end{eqnarray}
In our study, we mainly focus on the strong interaction regime where
the interaction energy $E_c$ dominates the tunneling coupling
$E_c/J>>1$ \cite{Lee,Chena,Kasztelan} and the high-frequency limit
$E_c/J>>\omega/J>>1$. Therefore, these terms on the right hand side of
Eq.~(\ref{dequationb}) are rapidly oscillating, and  make important
contributions only if $\varphi_{n}^m$ has a stationary phase at
certain instant determined by the following condition
\cite{Kayanuma, Ful}
\begin{eqnarray} \label{conditequation}
\frac{d\varphi_{n}^{m}}{dt}=E_c(2n+1)/2+\alpha t+m\omega,
\end{eqnarray}
thereby resulting in
\begin{equation}
\alpha t_n^m=-(n+1/2)E_c-m\omega.
\end{equation}
In the absence of the periodic driving field, $\delta_1=0$, the last
term $m\omega$ vanishes. We note that in the case of $\delta_1=0$,
the ground state undergoes $N$ LZ transitions at $\alpha
t_n=-(n+1/2)E_c$. So the LZ transitions at $t_n^0=t_n$ just
correspond to the usual many-body LZ process without periodic
driving. Naturally, a problem arises,  what is  the dynamics of
the system near $t_n^m$ with $m\neq 0$?

\begin{figure}[htb]
\begin{center}
\includegraphics[bb= 11 2 447 381, clip, scale=0.55]{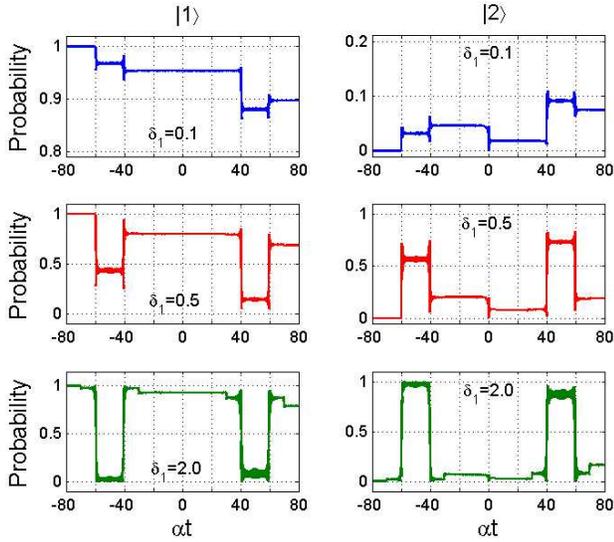}
\caption{(Color online) Occupation probability of the system in the
two lowest instantaneous eigenstates of the Hamiltonian
$H(t)_{\delta_1=0}$ for different driving amplitudes $\delta_1$ with
$N=2$,  $J=1$, $E_c=100$,  $\alpha=0.01$, and $\omega=10$.
Here the labels $|1\rangle$ and  $|2\rangle$ represent the two lowest instantaneous
eigenstates of the Hamiltonian $H(t)_{\delta_1=0}$.}
\label{fig2}
\end{center}
\end{figure}

To answer this problem, we first make a change of variable
$\tau=t-t_n^m$ near $t_n^m$, and then assume a high-frequency
driving $\omega/J>>1$ for simplifying our discussions. Because of
$\omega/J>>1$, we may retain the $m$-th term, and approximately
replace the Eq.~(\ref{dequationb}) by two first-order differential
equations \cite{Kayanuma, Ful}
\begin{eqnarray}
i\frac{dc_{n}}{d\tau}&=&-\widetilde{J}_{n,eff}^m e^{-i\phi_n^m}e^{-i\frac{\alpha \tau^2}{2}}c_{n+1},\\
i\frac{dc_{n+1}}{d\tau}&=&-\widetilde{J}_{n,eff}^{m}e^{i\phi_n^m}e^{i\frac{\alpha
\tau^2}{2}}c_{n},
\end{eqnarray}
where  $\phi_n^m=\alpha (t_n^m)^2/2+E_c(2n+1)t_n^m/2+m\omega t_n^m$.
Here $n$ takes from $-N/2$ to $N/2-1$. We note that  similar coupled
equation between $c_{n}$ and $c_{n-1}$ can also be obtained where
$n$ starts from $N/2$ to $-N/2+1$.

To clearly see the physics described by the above equations, we
make the transformation $c_n=\widetilde{c}_n\exp[-i\alpha \tau^2/4]$
and $c_{n+1}=\widetilde{c}_{n+1}\exp[i\alpha \tau^2/4]$, and get
\begin{eqnarray}
i\frac{d\widetilde{c}_{n}}{d\tau}&=&-
\frac{\alpha \tau}{2} \widetilde{c}_{n}-\widetilde{J}_{n,eff}^m e^{-i\phi_n^m}\widetilde{c}_{n+1},\\
i\frac{d\widetilde{c}_{n+1}}{d\tau}&=&\frac{\alpha \tau}{2}
\widetilde{c}_{n+1}-\widetilde{J}_{n,eff}^{m}e^{i\phi_n^m}\widetilde{c}_{n}.
\end{eqnarray}
These results  tell us that the change of the coefficients $c_n$ due
to a linear sweep  across $t=t_n^m$ is nothing but that of the
LZ-type level crossing in which the coupling is renormalized
effectively by a factor of the Bessel function.

Therefore, the LZ transitions at $\alpha t_n^0$ just correspond  to
the usual LZ transitions without a periodic driving field,  while the
sideband LZ transitions at $\alpha
 t_n^m$ with $m\neq 0$ arise from the periodic driving field of the
time-dependent bias.  In principle, the index $m$ can take arbitrary integer
values. However,  from the well-know fact that
$J_m(\delta_1/\omega)\rightarrow 0$ with $m\neq 0$ if
$\delta_1/\omega\rightarrow 0$ or $m\rightarrow\pm \infty$, it
follows that for the very small driving amplitude, the sideband  LZ
transitions  are so small that they are actually not visible. If the
driving amplitude is chosen suitably, they become important and
visible.

To confirm our analytical results with numerical simulations,  we
use the usual many-body LZ Hamiltonian $H(t)_{\delta_1=0}$ as a reference
system.  We solve numerically  the time-dependent Schr\"{o}dinger
equation $i\partial |\psi(t)\rangle/\partial t
=H(t)|\psi(t)\rangle$ starting from the ground state of
Hamiltonian (\ref{bhh}) with a large negative bias
$\delta(t=-T)\rightarrow-\infty$, and at the end of  the linear
sweep $\delta(t=T)\rightarrow \infty$ of the bias, and compute the
occupation probability of the system in
the lowest instantaneous eigenstates of the
Hamiltonian $H(t)_{\delta_1=0}$. In Fig. \ref{fig1}, we first
display the numerical results without the periodic driving field for
the small atom number $N=2$. Here the other parameters are given by
$J=1$ and $E_c=100$. This situation corresponds to the usual
many-body LZ problem. As is  expected, if the sweep rate $\alpha$ is
low enough, the system is still in the instantaneous ground state
during the linear sweep. However, the presence of the periodic
driving field modifies this physical picture. To only show the effect
of the periodic driving, we take a small sweep rate $\alpha=0.01$, for which
the evolution of the system without periodic driving field is
adiabatic. In
Fig.~\ref{fig2}, we display the occupation probability of the system
in the  two lowest instantaneous eigenstates of the Hamiltonian
$H(t)_{\delta_1=0}$ for different driving amplitudes with $N=2$, $J=1$,
$\alpha=0.01$, $\omega=10$ and $E_c=100$. We find from these
numerical results that the occupation probability displays  a series
of steplike changes at particular values of $\alpha t_n^m$ with
$m\neq 0$. In this situation, for the small driving amplitudes
$\delta_1=0.1$ and $\delta_1=0.5$,  the LZ transitions with $m=\pm
1$ are clearly visible, while for a larger driving amplitudes
$\delta_1=2.0$, the LZ transitions with $m=\pm 2$ become also
visible.

\section{Quantization of the periodic driving field}

In this section, we show how to understand the sideband transitions in our many-body LZ process by quantizing the periodic driving field. Usually, the quantization of a classical field is achieved by introducing a harmonic oscillator and then quantizing the harmonic oscillator. We introduce a hybrid quantum-classical system composed of a quantum subsystem and a classical harmonic oscillator, and then show that the coupling between the quantum subsystem and the classical harmonic oscillator can act as the periodic driving for the quantum subsystem.  Therefore, the quantization of the periodic driving field corresponds to the quantization of the classical harmonic oscillator in the hybrid quantum-classical system.

We consider a hybrid quantum-classical system
\begin{equation}
H_{hy}=H_q+H_{ho}+H_{q}^{ho},
\end{equation}
with
\begin{equation}
H_q=-2J S_x+\frac{E_c}{2}S_z^2+\alpha t S_z,
\end{equation}
for the quantum subsystem,
\begin{equation}
H_{ho}=\frac{P^2}{2M}+\frac{1}{2}M\omega^2Q^2
\end{equation}
for the classical harmonic oscillator, and
\begin{equation}
H_{q}^{ho}=k Q S_z,
\end{equation}
for the coupling between the quantum subsystem and the classical harmonic oscillator. Here $k$ is the coupling strength, $Q$ is the oscillator position, and $P$ is the oscillator momentum. If the mass $M$ is large enough, the harmonic oscillator will not be affected by the quantum subsystem and the quantum subsystem feels a periodic driving produced by the harmonic oscillator~\cite{makarov,makri}.

Through introducing the destruction and creation operators $b$ and $b^{\dagger}$ for the momentum $P$ and the position $Q$ of a harmonic oscillator,  the fully quantum model for the hybrid quantum-classical system may be written as
\begin{eqnarray} \label{qquationb}
H_{fq}&=&-2J S_x+\frac{E_c}{2}S_z^2+\alpha t S_z
+\lambda(b^{\dagger}+b)S_z+ \omega b^{\dagger}b .\nonumber \\
\end{eqnarray}
Here $\lambda=k/\sqrt{2}$ is the rescaled coupling strength. For the case of
$N=1$ and $E_c=0$, the resulting  model can be used to describe the LZ process in  a quantum two-level system coupled to a photon mode  \cite{nori}.
By employing a unitary transformation $U=e^{i\omega t b^{\dagger}b}$, the quantum Hamiltonian   (\ref{qquationb}) is equivalent to an effective Hamiltonian
\begin{eqnarray} \label{qquationbeff}
H_{fq}^{eff}&=&-2J S_x+\frac{E_c}{2}S_z^2+\alpha t S_z \nonumber \\
&&+\lambda(e^{i\omega t}b^{\dagger}+e^{-i\omega t}b)S_z .
\end{eqnarray}
Clearly, if the  harmonic oscillator stays in a  coherent state $\left|\beta\right\rangle$, the quantum subsystem just feels a periodic driving field induced by the coupling term $H_{q}^{ho}$ and so that it obeys,
\begin{eqnarray} \label{qquationcor}
H_{q}^{eff}&=&-2J S_x+\frac{E_c}{2}S_z^2+\alpha t S_z \nonumber \\
&&+\lambda\langle\beta|(e^{i\omega t}b^{\dagger}+e^{-i\omega t}b)|\beta\rangle S_z, \nonumber \\
&=&-2J S_x+\frac{E_c}{2}S_z^2+\alpha t S_z \nonumber \\
&&+2\lambda \beta \cos(\omega t)S_z.
\end{eqnarray}
Usually, we can assume that $\beta$ is a real number. Then the amplitude of the periodic driving $\delta_1$ is related to $2\lambda\beta$.

\begin{figure}[htb] \center
\begin{center}
\includegraphics[bb= 6 1 334 460, clip, scale=0.65]{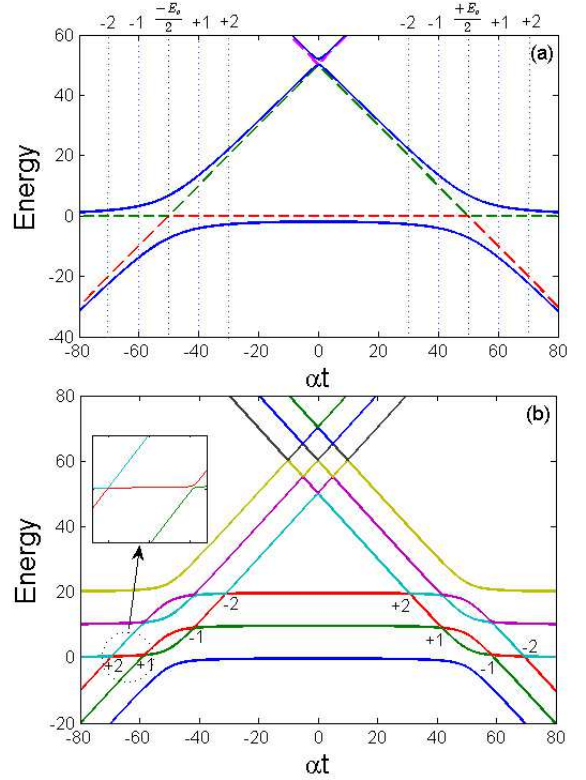}
\caption{(Color online) (a) Instantaneous energy spectra of the many-body LZ
Hamiltonian $H_{fq}$  for  $J=0$ (dashed lines) and $J=1$ (solid liens) with $\lambda=\omega=0$. Here the total atom number is $N=2$, and the interaction strength is $E_c=100$. (b) Instantaneous energy spectra of the many-body LZ
Hamiltonian $H_{fq}$ with  $\lambda=1$ and $\omega=10$. Here for simplicity, we only take a relatively small  basis set for the photon mode.  The labels  of -2,-1,+1 and +2 denote  the
avoided level-crossings at $\alpha t =
\pm E_c/2-m\omega$ with $m=-2,-1,+1,+2$. In (b), the inset shows the enlarged region near the avoided level-crossings labeled by $m=+2,+1$.
} \label{fig3}
\end{center}
\end{figure}

To compute  the energy spectrum of the  Hamiltonian $H_{fq}$ with
a fixed  $\alpha t$,  we use the basis $|n_b\rangle\otimes |n\rangle$
satisfying $b^{\dagger}b|n_b\rangle=n_b|n_b\rangle $ and
$S_z|n\rangle=n|n\rangle$. For a  small value of $\beta$, we may
take a finite basis set, $n_b=0,1,2$, and diagonalize the
Hamiltonian numerically. In Fig. \ref{fig3}, we display instantaneous energy spectra of the
Hamiltonian $H_{fq}$ as a function of $\alpha t$ for two different cases (a) $\lambda=\omega=0$ and (b) $\lambda=1$ and
$\omega=10$. In the two situations, the atom number is $N=2$, and
the other parameters are given as $J=1$ and $E_c=100$. For the
energy spectrum  without
the photon mode, the avoided level-crossings of the ground
state only appear around $\alpha t=\pm E_c/2$ and these avoided
level-crossings dominate the population transfer in the LZ
process, as shown in Fig. \ref{fig3} (a). For the energy spectrum
in  the presence of the photon mode, because of the many photon effects, the avoided
level-crossings can appear around $\alpha t=\pm Ec/2-m\omega$
with $m=0,\pm 1,\pm 2$. It is  observed that the gap around $\alpha t=\pm
Ec/2-m\omega$ with $m=\pm 1$ is lager than those for $m=\pm 2$, as illustrated in Fig. \ref{fig3}
(b). This results explain why the  sideband transitions  with $m=\pm 1$ are
observed in Fig. \ref{fig2}  for a relative small driving with
$\delta_1=0.1, 0.5$. When  the driving amplitude is increased to  a
larger value related to a larger $\beta$, we need to include more
photon numbers to compute the energy spectrum, and thus  more
sideband transitions may be observed. For example,  in Fig.
\ref{fig2}, the sideband transitions with $m=\pm 2$ are observed in
the case of $\delta_1=2$.

\section{Conclusions}

In summary, we have investigated the many-body LZ process in the
two-site Bose-Hubbard model where the energy bias between sites is
subject to a linear change with time and a periodic driving field. It
is revealed that the periodic driving can  modify the conditions
for   the many-body LZ transitions, and induce sideband LZ
transitions under certain parameter conditions. In the
high-frequency limit and strong interaction regime, we have
applied an analytical method to understanding these sideband LZ
transitions. In addition, by quantizing  the periodic driving field, we
introduce  the fully quantum mechanical model to understand the
sideband LZ transitions, in which the sideband transitions can be understood as the LZ transitions between states of different ``photon" numbers. Our results
show that
the periodic
driving field can provide an efficient way for controlling the many-body LZ
tunneling process.

Our results of sideband transitions in many-body LZ problem offer an alternative route to manipulating many-body quantum systems. 
With currently avaliable experimental techniques for
observing many-body LZ tunneling \cite{Chena}, it is possible to test
our theoretical predictions.
It is also possible to apply our analysis for treating the case of more complex driving fields, such as multi- frequency driving field, and the driving field with time-dependent amplitude.

\begin{acknowledgments}
H.-H. Zhong and Q.-T. Xie made equal contributions. This work is supported by the NBRPC under Grants No. 2012CB821305, the NNSFC under Grants No. 11374375, 11147021, 11375059, the Hunan Provincial Natural Science Foundation under Grant No. 12JJ4010, the  Scientific Research Fund of Hunan Provincial Education Department under Grant No. 13A058, and the Ph.D. Programs Foundation of Ministry of Education of China under Grant No. 20120171110022.
\end{acknowledgments}


\begin{references}
\bibitem{Landau}L. D. Landau, Phys. Z. Sowjetunion 2, 46 (1932).
\bibitem{Zener}C. Zener, Proc. R. Soc. London, Ser. A 137, 696 (1932).
\bibitem{Stueckelberg}E. C. G. St\"{u}eckelberg, Helv. Phys. Acta 5, 369 (1932).
\bibitem{Majorana}E. Majorana, Nuovo Cimento 9, 43 (1932).
\bibitem{mgong} Y. Qian, M. Gong, and C. Zhang, Phys. Rev. A 87, 013636 (2013).
\bibitem{longhi} S. Longhi and G. D. Valle, Phys. Rev. A 86, 043633 (2012).
 \bibitem{betthau} C. Betthausen, T. Dollinger, H. Saarikoski, V. Kolkovsky, G. Karczewski, T. Wojtowicz, K. Richter, and D. Weiss, Science 337, 324 (2012).
\bibitem{william} W. D. Oliver, Y. Yu, J. C. Lee, K. K. Berggren,
L. S. Levitov, and T. P. Orlando, Science 310, 1653 (2005).

\bibitem{Wu}B. Wu and Q. Niu, Phys. Rev. A 61, 023402 (2000).
\bibitem{Zobay}O. Zobay and B. M. Garraway, Phys. Rev. A 61, 033603 (2000).
\bibitem{Liu}J. Liu, L. Fu, B. Ou, S. Chen, D. Choi, B. Wu, and Q. Niu,
Phys. Rev. A 66, 023404 (2002).
\bibitem{Witthaut}D. Witthaut, E. M. Graefe, and H. J. Korsch, Phys. Rev. A 73, 063609 (2006).
\bibitem{Tomadin}A. Tomadin, R. Mannella, and S. Wimberger, Phys. Rev. A 77, 013606 (2008).
\bibitem{Lee}C. Lee, L.-B. Fu, and Y. S. Kivshar, Europhys. Lett. 81, 60006 (2008).
\bibitem{Smith}K. Smith-Mannschott, M. Chuchem, M. Hiller, T. Kottos, and
D. Cohen, Phys. Rev. Lett. 102, 230401 (2009).
\bibitem{korsch} F. Trimborn, D. Witthaut, V. Kegel, and H. J. Korsch, New J. Phys. 12,  053010 (2010).
\bibitem{lignier} A. Zenesini, H. Lignier, G. Tayebirad, J. Radogostowicz, D. Ciampini, R. Mannella, S. Wimberger, O. Morsch, and E. Arimondo, Phys. Rev. Lett. 103, 090403 (2009).
\bibitem{Kasztelan}C. Kasztelan, S. Trotzky, Y. Chen, I. Bloch, I. P. McCull\"{o}ch, U. Schollwock, and G.
Orso, Phys. Rev. Lett. 106, 155302 (2011).

\bibitem{Arimondo}M. Jona-Lasinio, O. Morsch, M. Cristiani, N. Malossi, J. H. M\"{u}ller, E. Courtade, M. Anderlini, and E.
Arimondo, Phys. Rev. Lett. 91, 230406 (2003).

\bibitem{Salger}T. Salger, C. Geckeler, S. Kling, and M. Weitz, Phys.
Rev. Lett. 99, 190405 (2007).
\bibitem{Kling}S. Kling, T. Salger, C. Grossert, and M. Weitz, Phys. Rev.
Lett. 105, 215301 (2010).


\bibitem{Chena}Y. Chen, S. Huber, S. Trotzky, I. Bloch, and E. Altman,
Nature Physics 7, 61 (2011).


\bibitem{grifoni98} M. Grifoni and P. H\"{a} nggi, Phys. Rep. 304, 229 (1998).




\bibitem{Kayanuma}Y. Kayanuma and Y. Mizumoto, Phys. Rev. A 62, R061401
(2000).
\bibitem{Ful} S. Duan, L. Fu, J. Liu, and X. Zhao, Phys. Lett. A 346,
315 (2005).
\bibitem{malossi}N. Malossi, M. G. Bason, M. Viteau, E. Arimondo, R. Mannella, O. Morsch, and D. Ciampini, Phys. Rev. A 87, 012116 (2013).
\bibitem{wubs} M. Wubs, K. Saito, S. Kohler, Y. Kayanuma, and P. H\"{a}nggi, New J. Phys. 7, 218 (2005).
\bibitem{qzhang} Q. Zhang, P. H\"{a}nggi, and J. Gong, New J. Phys. 10, 073008 (2008).
\bibitem{qzhang08} Q. Zhang, P. H\"{a}nggi, and J. Gong, Phys. Rev. A 77, 053607 (2008).

\bibitem{smerzi97} A. Smerzi, S. Fantoni, S. Giovanazzi, and S. R. Shenoy, Phys. Rev. Lett. 79, 4950 (1997).
\bibitem{corney} G. J. Milburn, J. Corney, E. M. Wright, and D. F. Walls, Phys.
Rev. A 55, 4318 (1997).
\bibitem{smerzi}A. S. Parkins and D. F. Walls, Phys. Rep. 303, 1 (1998).
\bibitem{martin} T. Jinasundera, C. Weiss, and M. Holthaus, Chemical Physics 322, 118 (2006).
\bibitem{gati2007} R. Gati and M. K. Oberthaler, J. Phys. B 40, R61 (2007).
\bibitem{Lee2012} C. Lee, J. Huang, H. Deng, H. Dai, and J. Xu, Front. Phys. 7, 109 (2012).



\bibitem{della} G. Della Valle, M. Ornigotti, E. Cianci, V. Foglietti, P. Laporta,
and S. Longhi, Phys. Rev. Lett. 98, 263601 (2007).
\bibitem{kartashov} Y. V. Kartashov and V. A. Vysloukh, Opt. Lett. 34, 3544 (2009).


\bibitem{makarov} D. E. Makarov, Phys. Rev. E 48, R4146 (1993).

\bibitem{makri}D. E. Makarov and N. Makri, Phys. Rev. E 52, 5863 (1995).
\bibitem{nori}Z. Sun, J. Ma, X. Wang, and F. Nori, Phys. Rev. A 86, 012107 (2012).
\end{references}
\end{document}